\documentclass[%
 preprint,
 amsmath,amssymb,
 aps,
]{revtex4-2}

\usepackage{graphicx}
\usepackage{dcolumn}
\usepackage{bm}
\usepackage{hyperref}

\begin{document}
\title{SLO/GO Degradation--Loss Sensitivity in Climate--Human System
Coupling}

\author{Sierra Cabrera}
\affiliation{Department of Physics, Engineering Physics,
and Astronomy, Queen's University, Kingston ON, K7L 3N6, Canada}
\author{Irina Babayan}
\affiliation{Department of Physics, Engineering Physics,
and Astronomy, Queen's University, Kingston ON, K7L 3N6, Canada}
\author{Hazhir Aliahmadi}
\affiliation{Department of Physics, Engineering Physics,
and Astronomy, Queen's University, Kingston ON, K7L 3N6, Canada}
\author{Dongmei Chen}
\affiliation{Department of Geography and Planning, Queen's University, Kingston
ON, K7L 3N6}
\author{Greg van Anders}
\affiliation{Department of Physics, Engineering Physics,
and Astronomy, Queen's University, Kingston ON, K7L 3N6, Canada}
\email{gva@queensu.ca}

\date{\today}

\begin{abstract}
  The potential of extreme environmental change driven by a destabilized climate
  system is an alarming prospect for humanity. But the intricate, subtle ways
  Earth's climate couples to social and economic systems raise the question of
  when more incremental climate change signals the need for alarm.  Questions
  about incremental sensitivity are particularly crucial for human systems that
  are organized by optimization. Optimization is most valuable in resolving
  complex interactions among multiple factors, however, those interactions can
  obscure coupling to underlying drivers such as environmental degradation.
  Here, using Multi-Objective Land Allocation as an example, we show that model
  features that are common across non-convex optimization problems drive
  hypersensitivities in climate-induced degradation--loss response. We show that
  catastrophic losses in human systems can occur well before catastrophic
  climate collapse. We find punctuated insensitive/hypersensitive
  degradation--loss response, which we trace to the contrasting effects of
  environmental degradation on subleading, local versus global optima (SLO/GO).
  We argue that the SLO/GO response we identify in land-allocation problems
  traces to features that are common across non-convex optimization problems
  more broadly. Given the broad range of human systems that rely on non-convex
  optimization, our results therefore suggest that substantial social and
  economic risks could be lurking in a broad range in human systems that are
  coupled to the environment, even in the absence of catastrophic changes to the
  environment itself.
\end{abstract}

\maketitle

\section{Introduction}
Climate change investigations \cite{scheffer_catastrophic_2001} indicate that
destabilizing Earth's climate system may drive extreme environmental shifts that
harbour catastrophic impacts on humanity. However, the coupling of human social
and economic organization to the climate is complex
\cite{turner_emergence_2007}, so the sensitivity of a given form of human
organization to climate change is unclear.\cite{turner_emergence_2007} For
example, will large-scale changes in how we organize our societies and economies
only stem from widespread environmental catastrophe?  Or, could the nature of
human--climate coupling drive large-scale changes in human
organization in response to minor climate change? Understanding
these questions is critical as data show Earth's climate continues to warm
\cite{IPCC-2022}.

Parameter sensitivity questions are crucial for land-use planning. Environmental changes,
such as increased flood risk \cite{FloodRisk}, reductions in precipitation
\cite{PrecipitationExtremes} or a decline in soil quality
\cite{seto_global_2012}, change the suitability of land for specific uses. These
changes have knock-on social and economic impacts, such as via insurance costs
\cite{zhengOptimisingLandUse2019} or crop yields \cite{IPCC-2022}, among other
factors. Moreover, even rapid land redevelopment or rehabilitation occurs on
significant timescales.\cite{China287Cities} If small changes in climate
conditions can trigger large-scale redistribution of optimal land-use patterns, the
mismatch in timescales between the redistribution trigger and the land-use
re-development response could create social and economic instability.
Understanding land-use decision making methods' resilience to climate change
therefore provides an important set of test cases for studying the relationship
between sensitivity and risk. In particular, land-use planning procedures can invoke
non-convex optimization techniques. Therefore, the way in which non-convexity drives
land-use planning responses to environmental degradation could signal the
existence of analogous effects in other systems of human organization.

\begin{figure*}
  \includegraphics[height=16cm]{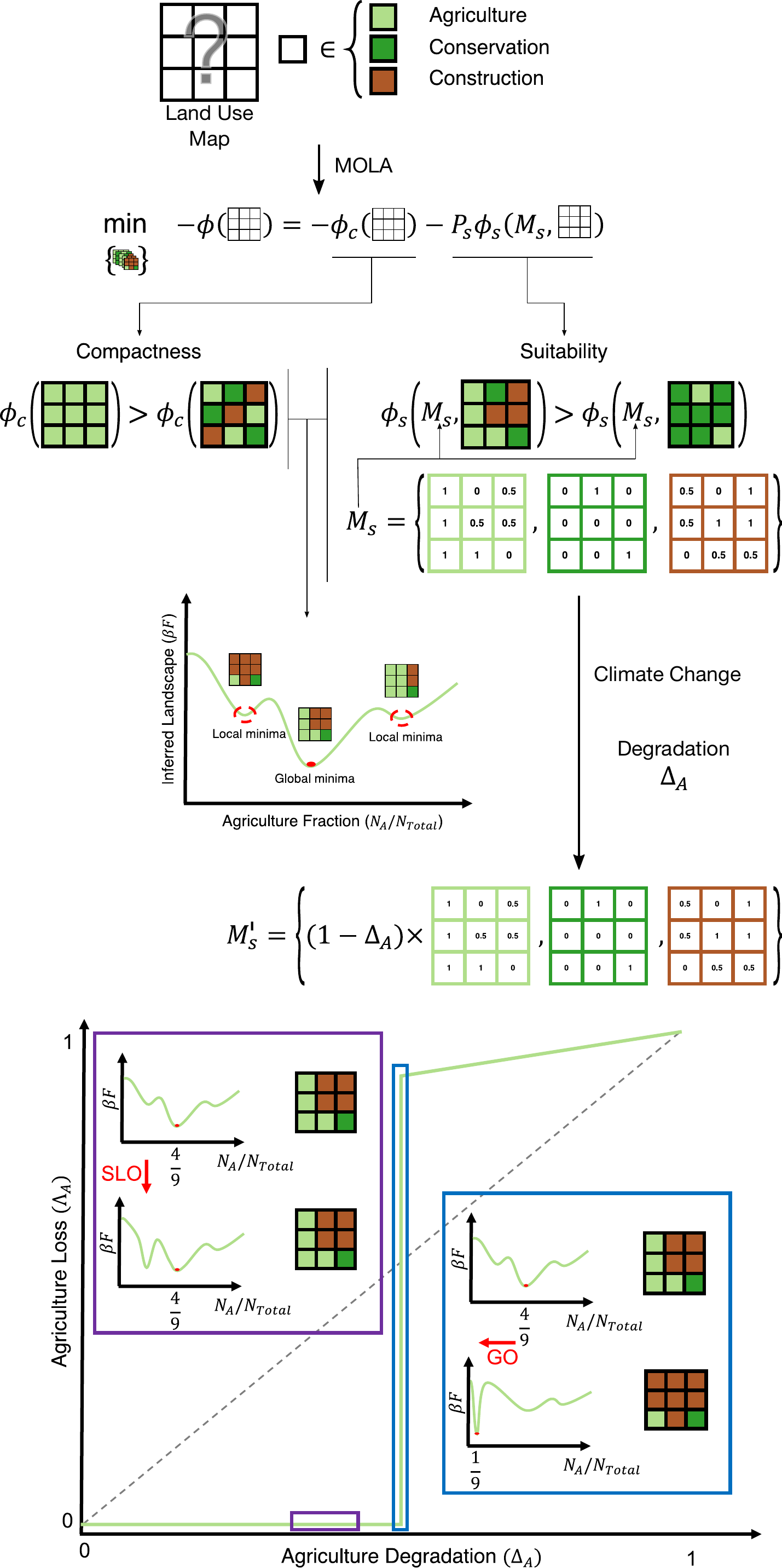}
  \caption{
    Schematic illustration of model of climate change-induced environmental
    degradation in land allocation. Multi-Objective Land Allocation (MOLA)
    supports land-use planning by casting the problem of determining land-use
    maps (here for simplicity we illustrate this via $3\times 3$ maps) as a
    weighted optimization problem. Typical planning criteria invoke notions of
    spatial compactness and land suitability, but optimal overall outcomes are
    typically not optimal for either component criterion, but instead involve
    complex trade-offs, and these trade-offs make it difficult to anticipate
    how planning would be affected by effects of climate change. We model this
    by incorporating potential future degradation in suitability for
    agricultural use. We compute the resulting loss of agricultural land under
    optimal scenarios in the face of degradation.
  }
  \label{fig:schematic}
\end{figure*}

Here, using a model \cite{SongChenMOLA} from the literature on geographic
information systems (GIS) informed land-use planning \cite{mola-review}, we show
that land-use planning exhibits punctuated hypersensitivity to climate change,
in which marginal land degradation can drive large-scale land reallocation. Via
a bespoke Markov Chain Monte Carlo (MCMC) sampling implementation in C++
\cite{connollyFlashWolff2023} we frame the results of $1.2\times 10^6$
statistically independent planning simulations in terms of a degradation--loss
($\Delta$--$\Lambda$) framework. Our computed $\Delta$--$\Lambda$ response
reveals that marginal changes in land suitability produce hypersensitive
land-use response. We find that the non-linear coupling between land allocation
and degradation results in the complete loss of specific land uses that occurs well
before complete degradation. We give a schematic representation of our approach
in Fig.\ \ref{fig:schematic}

To understand the origin and generality of these effects, we use brute force
methods to infer the parametric dependence of the landscapes of the underlying
non-convex optimization problem. We trace the origin of punctuated
insensitive/hypersensitive response to the non-uniform effect that environmental
degradation has on the underlying optimization landscape of the coupled human
system. It would be intuitive to expect that insensitive response occurs when
degradation has only a minor effect on the optimization landscape. We find,
instead, that degradation insensitivity can co-occur with large changes in
optimization landscapes if landscape changes are primarily concentrated among
sub-leading, local optima (SLO). In contrast to SLO-associated, insensitive
degradation response we find that hypersensitive response occurs when changes in
the optimization landscape affect global optima (GO).

Our analysis reveals two alarming outcomes: (i) that land-use planning models
exhibit hypersensitivity to marginal climate change, and (ii) that, in all test
cases, total loss greatly preceded total degradation. Additionally, our
analysis indicates that the punctuated insensitive/hypersensitive response we
observed was driven by systems with non-uniform, SLO/GO, parametric dependence
across the optimization landscape. The fact that this phenomenon can be traced
to such a generic feature of non-convex optimization is concerning because
non-convex optimization problems support social and economic organization in a
wide range of sectors such as airline networks,\cite{BiroliniAirNets} food
harvesting,\cite{FishOpt} energy distribution,\cite{ElecGasFlow} among others.
The generic mechanism that drives our findings thus signals that the SLO/GO induced
hypersensitivities we observed here could have alarming analogues in a broad
range of other areas of human activity. It is therefore vital to extend the
degradation--loss framework we presented here to identify and mitigate brewing
catastrophes that minor environmental change could induce in other sectors of
human social and economic activity.

\section{Results}
\subsection{Partial Degradation Drives Total Loss}
To determine the effect of climate change-induced degradation on land-use
distributions, we sampled land-use patterns from $1.2\times10^6$ statistically
independent MCMC simulations of a set of MOLA models
(see Methods). A MOLA model typically includes planning priorities of spatial
compactness and land suitability for a given set of uses, encoded as numerical
factors \cite{mola-review}. We used a reduction in the land-suitability factor
to model the effect of climate change via a degradation factor $\Delta_A$
assigned to agriculture that ranged between $0$ (no degradation) and $1$
(complete degradation). We extracted optimal land-use patterns from sampled
configurations, and computed agriculture loss $\Lambda_A$, the fractional
reduction in land-use for agriculture,
between $0$ (no loss compared to coverage at $\Delta_A=0$) and $1$ (complete
loss compared to coverage at $\Delta_A=0$).

\begin{figure*}
  \includegraphics[width=1.0\textwidth]{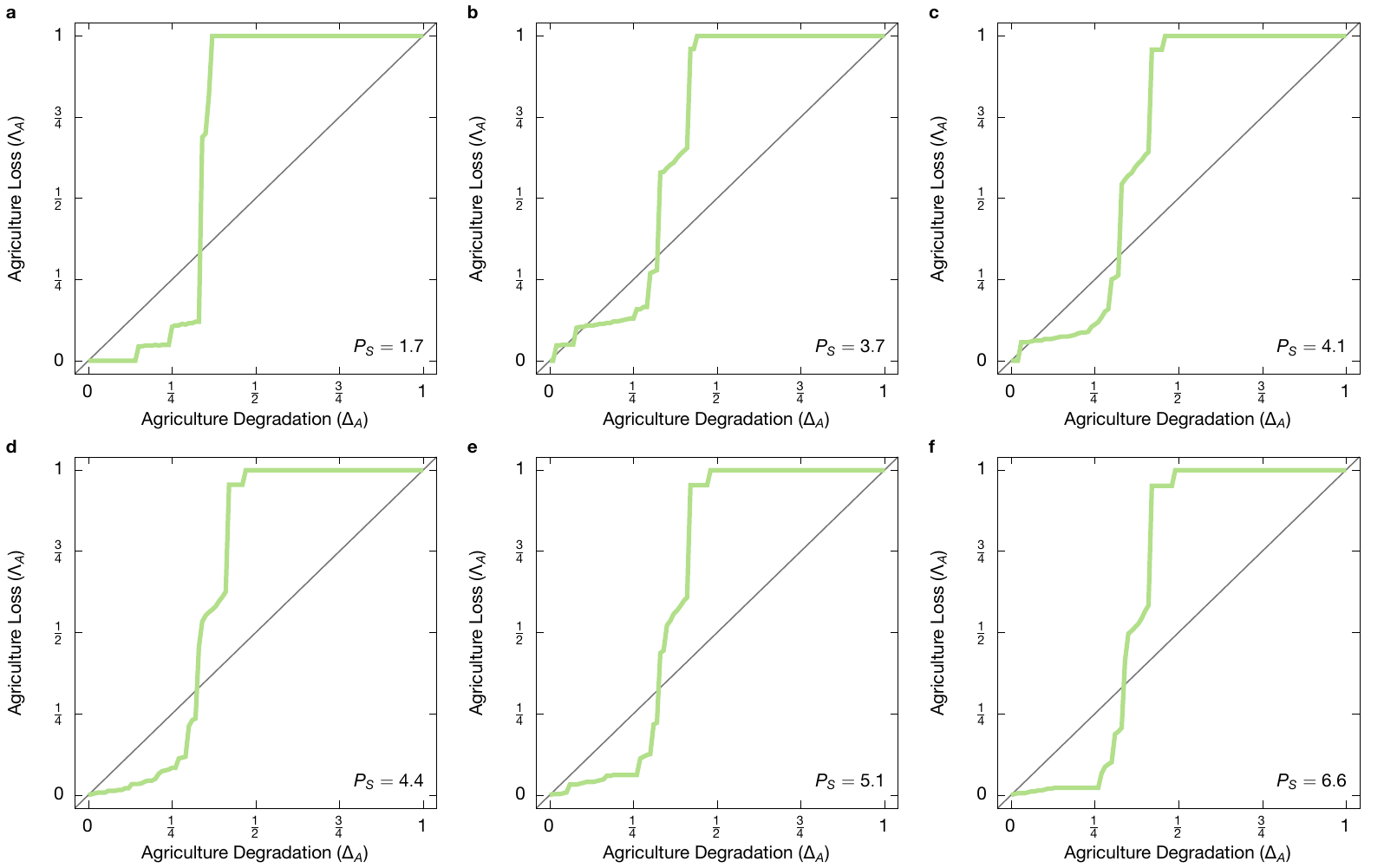}
  \caption{
    Degradation--Loss ($\Delta$--$\Lambda$) response of climate-coupled
    Multi-Objective Land Allocation (MOLA) shows that moderate degradation
    induces total loss and that marginal degradation produces punctuated
    insensitive/hypersensitive response. Panels a-f show $\Delta$--$\Lambda$
    response in six compactness--suitability trade-off regimes. Results are for
    uniform relative agricultural degradation $\Delta_A$ where 0 represents no
    degradation, and 1 represents total degradation. Vertical axes represent
    relative land-use loss $\Lambda_A$.  $\Lambda_A$ is normalized so that
    baseline land use coverage is taken relative to zero degradation (i.e.\
    $\Lambda_A(\Delta_A=0)\equiv 0$) and total loss corresponds to
    $\Lambda_A=1$. $P_S$ values (inset) parameterize the relative weight of
    suitability relative to compactness in the model.  $P_S$ values are selected
    to fall in compactness/suitability trade-off regimes identified in a prior,
    zero degradation investigation. In all cases $\Lambda_A=1$ for
    $\Delta_A<\tfrac{1}{2}$ indicating that moderate environmental degradation
    induces total land use loss. Additionally, the jagged, step-wise form of the
    response curves indicates that all cases exhibit punctuated
    insensitive/hypersensitive $\Delta$--$\Lambda$ response.
  }
  \label{Fig-LD_composite}
\end{figure*}

MOLA outcomes depend strongly on the relative prioritization of planning
objectives. A prior investigation of the current model without degradation
\cite{Flashpoints} showed that significant compactness--suitability trade-offs
occur when the relative priority, parameterized via a so-called suitability
pressure $P_S$, falls in the range $(0,8)$. We selected six different values of
$P_S$ to ensure that our analysis addresses distinct trade-off regimes
previously found in Ref.\ \cite{Flashpoints} in the absence of degradation.
$\Delta$--$\Lambda$ response for six trade-off regimes is shown in Fig.\
\ref{Fig-LD_composite}.

$\Delta$--$\Lambda$ response in Fig.\ \ref{Fig-LD_composite} indicates that in
all studied cases, the onset of total loss $\Lambda_A=1$, occurs well before the
advent of total degradation, $\Delta_A=1$. Indeed, we find that in all cases we
studied $\Lambda_A\to1$ for $\Delta_A<\tfrac{1}{2}$. The finding that partial
degradation drives total loss is concerning because it indicates that as climate
change-induced environmental degradation feeds into a human system organized by
non-convex optimization, even moderate degradation can lead to potentially
catastrophic changes in land-use patterns.

\subsection{Punctuated Insensitive/Hypersensitive Degradation--Loss
Response}
We next examine the degradation--loss ($\Delta$--$\Lambda$) response before the
onset of total loss.  This is also shown in Fig.\ \ref{Fig-LD_composite} and
corresponds to a portion of the $\Delta$--$\Lambda$ response curve that falls to
the left of the onset of total loss. In all six compactness--suitability
trade-off regimes we studied, we find that the $\Delta$--$\Lambda$ response is
characterized by a series of step-wise increases in loss. This step-wise
form signals that agricultural land-use loss ($\Lambda_A$) has a punctuated
insensitive\slash hypersensitive dependence on the degree of agriculture
degradation ($\Delta_A$).

A better understanding of how $\Delta$--$\Lambda$ data support this interpretation
requires comparing with general expectation. In general $\Delta$--$\Lambda$
response should pass through the points $(0,0)$ and $(1,1)$, i.e., we define
loss relative to a baseline land-use fraction at zero degradation, and we expect
that complete degradation will lead to complete loss. The fragility or
resilience of the $\Delta$--$\Lambda$ response is determined by how the system
responds between these extremes. The grey line in each plot gives the response
of a hypothetical system that exhibits a linear response that is commensurate with
expectations for zero and complete loss. With this framing, it is possible to
use $\Delta$--$\Lambda$ plots to identify both overall- and marginal sensitivity.
Low overall sensitivity can be identified as the degree to which $\Delta$--$\Lambda$
response falls in the range $\Lambda<\Delta$ (i.e., overall relative degradation
exceeds overall relative loss). Conversely, high overall sensitivity can be
identified as the degree to which $\Delta$--$\Lambda$ response falls in the
range $\Lambda>\Delta$ (i.e., overall relative loss exceeds overall relative
degradation). Low sensitivity $\Lambda<\Delta$ indicates resilient
climate--human system coupling that suppresses degradation effects. In contrast
$\Lambda>\Delta$ indicates fragile climate--human system coupling that
exacerbates degradation effects. Similarly, for marginal $\Delta$--$\Lambda$
sensitivity, ``flat'' (i.e.\ $d\Lambda/d\Delta\approx 0$) marginal response signals
low sensitivity, which indicates marginal stability or resilience. In contrast
``vertical'' (i.e.\ $d\Lambda/d\Delta\to\infty$) marginal response signals
hypersensitivity, which indicates marginal instability or fragility. Moderate
sensitivity falls between these extremes.

Using the above lens for interpreting sensitivity, what we observed in Fig.\
\ref{Fig-LD_composite} is not indicative of moderation. Instead, in all cases we
observe that $\Delta$--$\Lambda$ response exhibits periods of low sensitivity,
in which increasing $\Delta_A$ results in little to no change in $\Lambda_A$,
punctuated by instances of hypersensitive response in which small changes in
$\Delta_A$ produce very large changes in $\Lambda_A$. Our results indicate that
the overall response appears resilient up to $\Delta_A\sim0.3$ (with some
$P_S$-dependent variation), after which there are a series of hypersensitivities
that render the overall response largely fragile.

Taken together, our results indicate that marginal increases in degradation can
induce large-scale losses, and that complete collapse of the land-use planning
system can occur long before the complete degradation of the underlying climate
system it is coupled to.

\subsection{SLO/GO Landscape Rearrangement Drives Punctuated Hypersensitive
Response}
\begin{figure*}
  \includegraphics[width=1.0\textwidth]{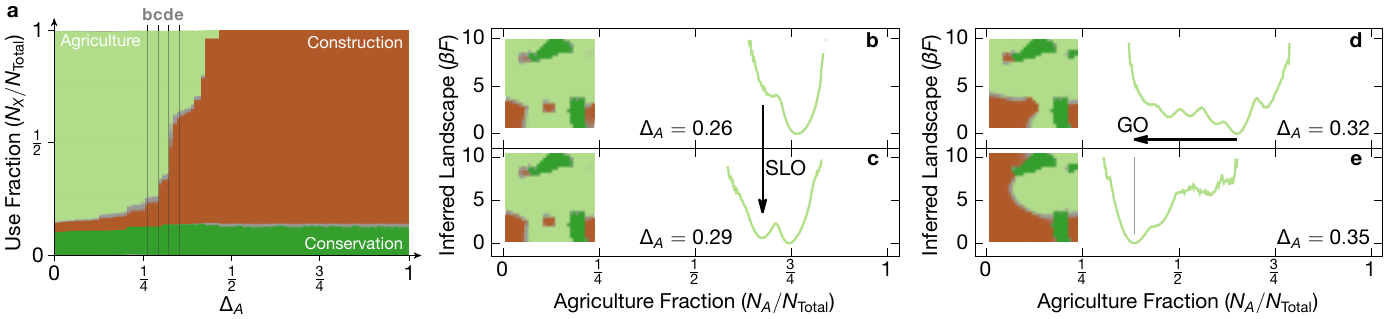}
  \caption{
    Changes in land-use patterns associated with hypersensitive
    degradation--loss ($\Delta$--$\Lambda$) response are driven by contrasts
    between subleading local optima (SLO) versus global optima (GO)
    rearrangements in underlying optimization landscapes. Panel \textbf{a} shows land-use
    distribution by degree of degradation shows punctuated
    insensitive/hypersensitive response in use patterns following
    $\Delta$--$\Lambda$ response shown in Fig.\ 1 (data are shown here for
    $P_S=4.4$, all other studied values of $P_S$ are shown in SI). This
    observed response arises from contrasting SLO/GO reorganization in the
    underlying landscapes. Panels \textbf{b} and \textbf{c} show use maps (inset) and 
    inferred optimization landscapes at degradation amounts separated by 3\% (b: 
$\Delta_A=0.26$; c: $\Delta_A=0.29$). The landscape plots show that this marginal 
    increase in degradation results in minor changes in optimal land use patterns. 
    However, this apparent insensitivity masks significant SLO rearrangement in 
    inferred underlying optimization landscapes (see Methods). Panels \textbf{d} and 
    \textbf{e} show use maps (inset) and inferred optimization landscapes at degradation 
    amounts separated by 3\% (d: $\Delta_A=0.32$; e: $\Delta_A=0.35$). Landscape plots
    show that this minor change in degradation affects underlying landscape's GO
    organization, which is echoed by macroscopic rearrangement of the spatial
    land-use distribution.
  }
  \label{Fig-why-figure}
\end{figure*}

To understand the scope of these observations, it is instructive to identify their
origin, which must trace back to the structure of the landscape of the underlying
optimization problem. Fig.\ \ref{Fig-why-figure}a shows the optimal land use
fraction for all land uses as a function of $\Delta_A$ at fixed $P_S=4.4$ (see
SI for other suitability pressures). To understand the origin of the difference
between insensitive and hypersensitive response, we inferred underlying
optimization landscapes from MCMC sampled allocation patterns. Key differences
between insensitive and hypersensitive $\Delta$--$\Lambda$ response can be seen
by comparing optimization landscapes for four values of degradation, $\Delta_A$,
spaced by increments of 3\% in Fig.\ \ref{Fig-why-figure}b,c,d,e. 

Between degradation $\Delta_A=0.26$ (Fig.\ \ref{Fig-why-figure}b) and $0.29$
(Fig.\ \ref{Fig-why-figure}c) there is little change in the land-use fraction
(Fig.\ \ref{Fig-why-figure}a) and there is little change in the spatial
distribution of land uses (inset maps in Fig.\ \ref{Fig-why-figure}b,c).
However, this marginal degradation-insensitivity in the optimal land pattern
does not mean that there are no changes in the underlying optimization
landscape. In fact, as a comparison between Fig.\ \ref{Fig-why-figure}b and c
shows, the optimization landscape reconfigures significantly between
$\Delta_A=0.26$ and $0.29$, but the primary effect is on the sub-leading, local
optimum (SLO).

As in Fig.\ \ref{Fig-why-figure}b and c, the marginal increase in degradation
between panels Fig.\ \ref{Fig-why-figure}d and e is also 3\%. However, the
effect that increased degradation has on the land use fraction (Fig.\
\ref{Fig-why-figure}a) and the spatial distribution land uses is dramatic.
Comparing the underlying optimization landscapes at $\Delta_A=0.32$ (Fig.\
\ref{Fig-why-figure}d) with that of $0.35$ (Fig.\ \ref{Fig-why-figure}e) reveals
that a relatively small increase in degradation affects the global optimum (GO),
resulting in a reduction in the agricultural use fraction from $\approx 0.65$.
to $\approx 0.38$. Normalized to a zero-degradation baseline, in this case a 3\%
change in degradation results in a relative loss of 32\%.

Therefore, the punctuated nature of the response is caused by underlying optimization landscape rearrangement,
with changes to SLO driving periods of insensitivity, and changes to GO producing hypersensitive response.

\section{Discussion}
We modelled the effect of climate-change induced environmental degradation on
land-use loss via multi-objective land allocation. Our degradation--loss
analysis yielded two key outcomes. We found that the nature of the coupling of
environmental degradation to a non-convex optimization model of land-use
planning resulted in the onset of complete loss of land use well in advance of
complete degradation of land-use suitability. Furthermore, we found that marginal
increases at intermediate amounts of degradation produced punctuated
insensitive/hypersensitive response in which increased degradation generally
led to one of two outcomes: almost no losses in land use, or large amounts of loss. Moreover, we
traced insensitive/hypersensitive response to reorganizations in the underlying
optimization landscape that primarily affected the configuration of subleading,
local optima (SLO) or the configuration of global optima (GO).

The SLO/GO-induced, punctuated insensitive/hypersensitive response we found here
results from the differential effects of degradation on the form of the
underlying landscape that originate, in turn, from the inhomogeneity of the land
itself. Intrinsic spatial inhomogeneity gives rise to a set of local minima that
respond unevenly to progressive environmental degradation. More formally, we
observed that punctuated insensitive/hypersensitive response coincided with an
optimization problem that had an optimization landscape with multiple local
minima, and that these local minima had different parametric dependence on
degradation. However, optimization landscapes with multiple local minima that
have different parametric dependence are ubiquitous across non-convex
optimization problems. The fact that a generic feature of non-convex
optimization produces the SLO/GO behaviour that drives the punctuated
insensitive/hypersensitive response we identified here suggests that a similar
response could arise generically in situations where human systems are coupled to
climate-induced environmental degradation.

The potential for the effects we observed to have analogues in other systems is concerning. 
Our investigation revealed that the effects of degradation can be easily masked in studies of human systems
organized by optimization if the parametric dependence of
the underlying landscape is not determined. The effects of degradation could then accumulate
among subleading local minima, unnoticed, and drive
subsequent hypersensitive response, seemingly without warning. I.e., in
optimization settings, identifying weak marginal response to small amounts of
environmental degradation provides no guarantee of subsequent stability.

Questions of stability are salient for, but not limited to, land-use. Triggering small, marginal environmental degradation could prompt response in
the form of redevelopment and rehabilitation, human migration, or commercial and
industrial relocation, all which occur over much longer timescales than the degradation. Other systems of
social or economic organization could be imperilled by trigger/response
timescale mismatches arising from SLO/GO-induced degradation--loss
hypersensitivity if their dynamic response is
limited by factors such as relocation, construction, manufacturing, procurement,
deployment, or distribution. Worryingly, most social and economic systems that
have a physical footprint are affected by multiple factors on that list. This
suggests that SLO/GO features of optimization-based organization could prompt
large-scale social and economic destabilization in many sectors even without
large-scale destabilization of the climate itself. Given the importance of
mitigating such instabilities, it would be valuable to extend the analysis given here, of 
elucidating the structure of the parametric dependence of the optimization landscape, to anticipate similar
SLO/GO-induced instabilities in other human systems.

\section{Methods}
\subsection{Land Degradation Model}
We employ a MOLA model from Ref.\ \cite{Song_Chen_data} that represents a 9
km$^2$ square area in the Xin’andu township of Dongxihu District, Wuhan, China
\cite{Song_Chen_data} as a 30 by 30 grid. Each grid cell represents a
parcel of land that can have one of three land use types: agriculture,
construction, or conservation.

Land-use planning in such a setting can be informed by the use of weighted,
multi-objective optimization based approaches to allocation. The model considered in Ref.\
\cite{SongChenMOLA} included spatial compactness and land suitability, two of
the most widely used criteria \cite{mola-review} in optimal land-use allocation. 

We aim to minimize an objective function that expresses the weighted combination
of planning criteria, as suggested in Ref.\ \cite{SongChenMOLA}, 
\begin{equation}
\label{eqn:Hamiltonian}
\phi = -P_C \phi_{C} - P_S \phi_{S} \; .
\end{equation}
In this equation, $P_C$ and $P_S$ are weights assigned to the objectives for
compactness, $\phi_{C}$, and suitability, $\phi_{S}$ (explicit forms for
$\phi_{C,S}$ are given below). $P_{C,S}$ can be interpreted as analogous to
physical pressure \cite{systemphys}. The allocation's characteristics are
influenced by the relative values of $P_C$ and $P_S$. A predominance of one
criterion over the other is seen when there is a significant disparity between
$P_C \phi_{C}$ and $P_S \phi_{S}$. Conversely, a trade-off between the two
criteria is achieved when $P_C \phi_{C}$ and $P_S \phi_{S}$ are of similar
magnitudes.

The compactness criterion $\phi_{C}$ is given by
\begin{equation}
\label{eqn:Compactness}
\phi_{c}=\sum_{k=0}^{K}\sum_{i=1}^{N}\sum_{j=1}^{M} c_{ijk} x_{ijk}  \; .
\end{equation}
Here, $M$ and $N$ index the map's columns and rows, while $K$ denotes the
number of distinct land-use types. The variable $x_{ijk}$ indicates whether the
parcel at coordinates $(i,j)$ is designated for land-use type $k$, in which case
$x_{ijk}=1$, otherwise it is $0$. The coefficient $c_{ijk}$ represents the
number of matching neighbours for the parcel at $(i,j)$ and is defined as
\begin{equation}
c_{ijk} = x_{i+1 j+1 k} + x_{i j+1 k} + x_{i+1 j k} + x_{i-1 j k} + x_{i j-1 k}+
x_{i-1 j-1 k} + x_{i-1 j+1 k} + x_{i+1 j-1 k} \; .
\end{equation}
The suitability criterion $\phi_{S}$ is given by
\begin{equation}
\label{eqn:Suitability}
\phi_{S}=\sum_{k=0}^{K}\sum_{i=1}^{N}\sum_{j=1}^{M} s_{ijk} x_{ijk} ; ,
\end{equation}
The suitability coefficient, $s_{ijk}$, quantifies the suitability of the parcel
at $(i,j)$ for a specific land-use type $k$. We consider a $30 \times 30$ map as
per Ref.\ \cite{SongChenMOLA}, with $M=N=30$, and three land-use types:
agriculture, construction, and conservation, represented by $k=0,1,2$.
Suitability data $s_{ijk}$ are detailed in the Supplementary Information (SI).

We investigated the trade-off regimes identified in Ref.\ \cite{Flashpoints} in
a modified form of the model where we modelled climate-induced environmental
degradation as a spatially-uniform scaling by a coefficient $\Delta_A$ of the
land-use suitability for agriculture. We parameterized degradation between
$\Delta_A=0$, zero degradation (baseline case), and $\Delta_A=1$ (total
degradation). Mathematically, the suitability coefficient for agriculture,
$k=0$, is scaled by $1-\Delta_A$ factor as 
\begin{equation}
\label{eqn:Degrade_H}
s'_{ij0} = (1-\Delta_A) s_{ij0} \; ,
\end{equation}
where $s'_{ijk} = s_{ijk}$ for all other $k$. 

A prior investigation of compactness/suitability trade-offs in this model
\cite{Flashpoints} identified that land-use patterns could be grouped into six
different trade-off regimes according to the relative weight of the design
objectives. Ref.\ \cite{Flashpoints} found that the key trade-off regime could
be found in terms of the suitability pressure  $0<P_S<8$ in units that fix
$P_C=1$.

Data shown in Figs.\ \ref{Fig-LD_composite} and \ref{Fig-why-figure} investigate
the sensitivity of loss response to the variation of degradation, $\Delta_A$, at
each trade-off regime. To quantify the relative loss, we use the number of
agricultural parcels and normalize them as
\begin{equation}
\label{eqn:Loss_A}
\Lambda_A = \frac{N_A(0) - N_A(\Delta)}{N_A(0)} \; ,
\end{equation}
where $N_A$ is the number of agricultural parcels.

\subsection{Markov Chain Monte Carlo Sampling}
For a suitability pressure, $P_S$, within distinct, zero-degradation trade-off
regimes, we computed optimal land use patterns via a cluster Markov Chain Monte
Carlo (MCMC) code written in C++. Our MCMC code implements the ghost-site Wolff
algorithm.\cite{GhostWolff} The code is available open source at Ref.\
\cite{connollyFlashWolff2023}.

Simple arguments based on detailed balance, see e.g.\
\cite{LandauBinderMC}, dictate that for any choice of $P_S$ and $\Delta_A$, the
most frequently sampled configuration is the one that is optimal. Because we
expect that the underlying optimization is non-convex, we used brute force
replication to correct for sampling errors. Overall, we ran more than
$1.2\times 10^6$ statistically independent simulations. The results of these
simulations are aggregated in Figs.\ \ref{Fig-LD_composite} and
\ref{Fig-why-figure}.

\subsection{Inferred Landscapes}
We inferred underlying optimization landscapes by appropriating free energy
methods from physics. In particular, we computed the Landau free energy
\cite{goldenfeld} using brute force accumulation following Ref.\
\cite{Flashpoints} analogously to methods used in systems of particles
\cite{entint}. 

By leveraging general arguments that relate optimization and statistical physics
\cite{DesignFilter}, it is possible to determine that the probability of $p$
sampling a given state that satisfies the design objective with a value of
$\phi$ is given by
\begin{equation}
  p = e^{-\phi/T} \; ,
  \label{eqn:prob}
\end{equation}
where $T$ is a temperature parameter used in MCMC sampling. Hence, by simply
accumulating samples, it is possible to reconstruct the solution space landscape
by computing $-\log(p)$ of the sampled distribution at fixed $T$. Because the
solution space space is $3^{30\times 30}$ extracting meaningful understanding
from individual land-use maps is impractical, however it is useful to aggregate
maps according to the total fraction of land-use of each type. Our inferred
landscapes report these aggregates.

\begin{acknowledgments}
We thank Jue Wang for helpful discussion. We acknowledge the support of the
Natural Sciences and Engineering Research Council of Canada (NSERC) grants
RGPIN-2019-05655, DGECR-2019-00469, and RGPIN-2019-05773. Computations were
performed on resources and with support provided by the Centre for Advanced
Computing (CAC) at Queen's University in Kingston, Ontario. The CAC is funded
by: the Canada Foundation for Innovation, the Government of Ontario, and Queen's
University. GvA thanks the hospitality of the Kavli Institute for Theoretical
Physics, where part of this work was done. This research was supported in part
by the National Science Foundation under Grant No.\ NSF PHY-1748958.
\end{acknowledgments}

\appendix
\section{Competing Interests}
The authors have no competing interests to declare.

\section{Data Availability}
Data from this investigation are available for download at Ref.\
\cite{SLOGOData}.

\section{Code Availability}
C++/Python source code used for this investigation is available for download at
Ref.\ \cite{connollyFlashWolff2023}.

\section{Author Contributions}
GvA initiated research. GvA, HA, and DC designed research. SC contributed
original code. SC and IB performed simulations. SC, IB, HA, and GvA analyzed
data. HA and GvA supervised research. SC, HA, DC, and GvA wrote the paper. All
authors contributed comments and edits on the manuscript.

\section{Supplementary Information}
Supplementary Figs. S1-S5.

\includegraphics[width=1.0\textwidth]{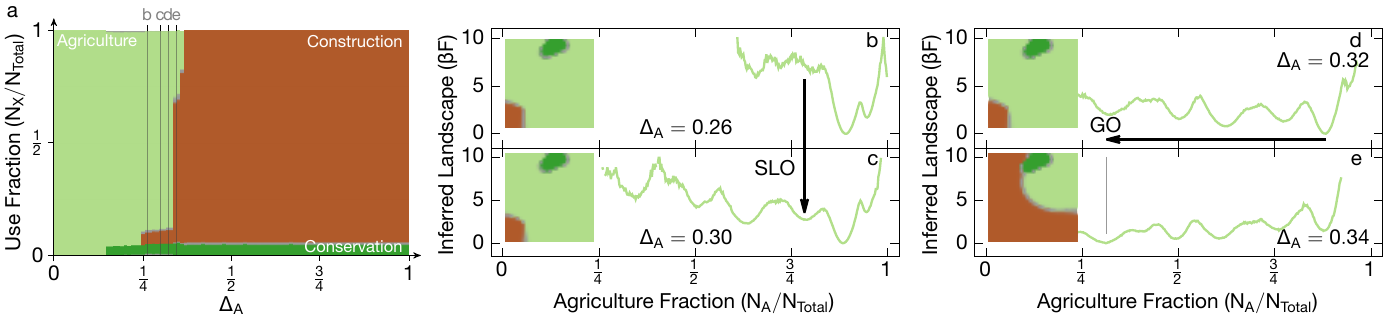}

\includegraphics[width=1.0\textwidth]{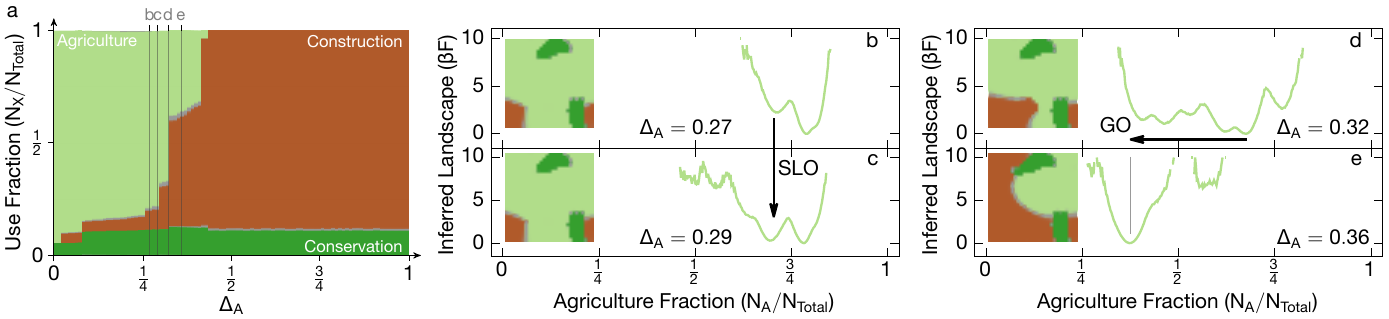}

\includegraphics[width=1.0\textwidth]{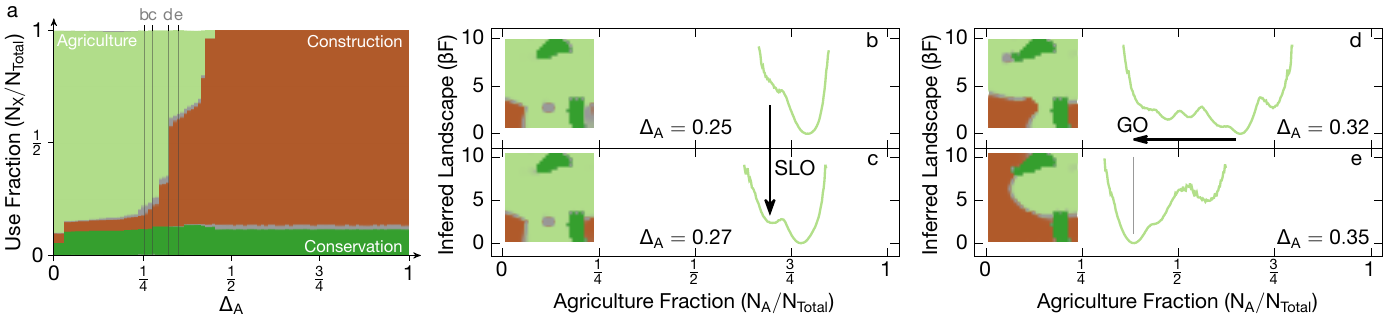}

\includegraphics[width=1.0\textwidth]{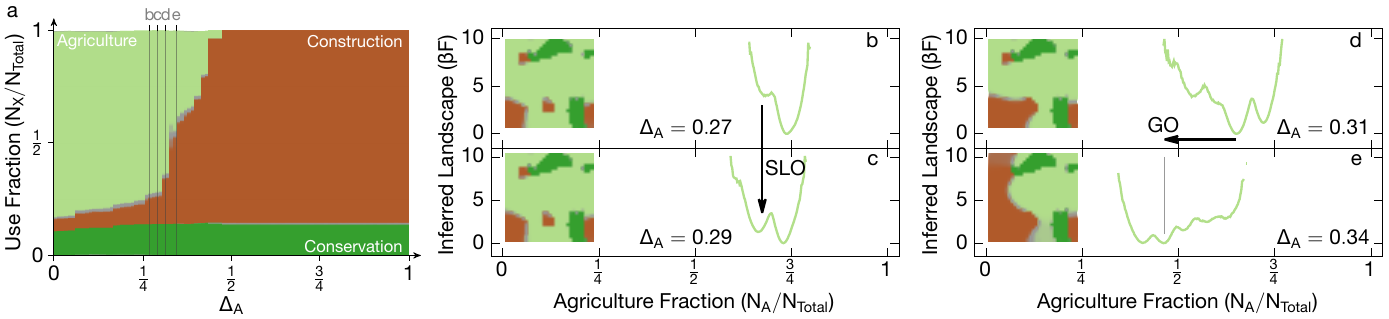}

\includegraphics[width=1.0\textwidth]{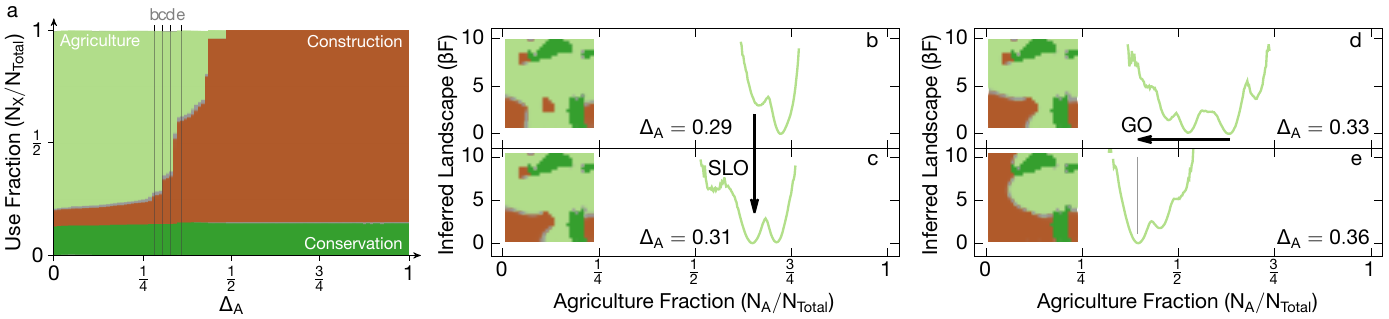}

\end{document}